# The Loeb Scale: Astronomical Classification of Interstellar Objects

Omer Eldadi[1], Gershon Tenenbaum[1] and Abraham Loeb[2]

1. B. Ivcher School of Psychology, Reichman University, Herzliya, Israel

2. Department of Astronomy, Harvard University, Cambridge, MA, USA

**Corresponding Address**:

Omer Eldadi

B.Ivcher School of Psychology

Reichman University

The University 8, Herzliya, Israel

Email: Omereldadi@gmail.com

## Abstract

The Vera C. Rubin Observatory is expected to increase interstellar object (ISO) detections from a few over the past decade to potentially one per few months, demanding a systematic classification scheme. We present the Loeb Scale, formally the Interstellar Object Significance Scale (IOSS), a 0-10 classification system extending the proven Torino Scale framework, to address ISOs' unique anomalies, including potential technosignatures. The scale provides quantitative thresholds for natural phenomena (Levels 0-3) and graduated protocols for increasingly anomalous characteristics (Levels 4-7), with Levels 8-10 reserved for confirmed artificial origin. Each level specifies observable criteria and response protocols. We demonstrate the scale's application using 1I/'Oumuamua (Level 4), 2I/Borisov (Level 0), and 3I/ATLAS (Level 4) as test cases. The Loeb Scale provides the astronomical community with a standardized framework for consistent, evidence-based and dynamic evaluation while maintaining scientific rigor across the full spectrum of possibilities as we enter an era of routine ISO encounters.

**Introduction**

The Vera C. Rubin Observatory's Legacy Survey of Space and Time (LSST) will transform interstellar object (ISO) detection from rare serendipity to routine observation, with projections indicating an increase from the few detections of 1I/ʻOumuamua, 2I/Borisov and 3I/ATLAS over the past decade to potentially one new ISO per few months (Dorsey et al., 2025; Hoover et al., 2022; Siraj & Loeb, 2022). This increase in detection rate by nearly two orders-of-magnitude demands immediate reconsideration of how the scientific community prepares for and responds to these cosmic objects. While most ISOs will likely prove to be natural phenomena, the sheer volume of detections raises profound questions about our readiness for discoveries that could challenge fundamental assumptions about our place in the universe.

The regular arrival of objects from other star systems is now a predictable feature of our cosmic environment, offering unprecedented scientific opportunities ranging from novel material compositions and properties to potential biosignatures or even technosignatures (Desch & Jackson, 2021; Hein et al., 2022; Lingam & Loeb, 2019). Each ISO that passes through our solar system without a comprehensive examination of its anomalies, as occurred with 1I/ʻOumuamua, 2I/Borisov and possibly with 3I/ATLAS, represents an irretrievable loss of scientific knowledge. This unprecedented situation requires us to examine our current preparedness and the development of a conceptual framework capable of systematically evaluating ISOs as their detection rate increases. This article presents the Loeb Scale[1] (the Interstellar Object Significance Scale; IOSS), a 0-10 classification system that extends the proven Torino Scale framework to address the unique anomalies of some ISOs, including explicit protocols for evaluating potential technosignatures.

---

[1] The idea for this scale was originally suggested in an essay written by A. Loeb in July 2025, https://avi-loeb.medium.com/the-visionary-letter-from-congresswoman-anna-paulina-luna-to-nasa-regarding-3i-atlas-ddb56dce69f0

**Institutional Barriers to Comprehensive ISO Classification**

The scientific response to 1I/'Oumuamua revealed a critical gap in scientific assessment frameworks. Despite its anomalous non-gravitational acceleration (Micheli et al., 2018) and unusual aspect ratio that defied easy explanation (Drahus et al., 2018; Meech et al., 2017), proposals to systematically evaluate artificial origin hypotheses met with resistance that exceeded normal scientific skepticism (Bialy & Loeb, 2018; Curran, 2021; Lineweaver, 2022; Loeb, 2022; Zuckerman, 2022). This response reflects broader institutional barriers: technosignature research remains marginalized within astronomy, receiving nearly no federal funding despite its potential transformative impact (Astro2020, 2023). These constraints create a troubling paradox: current classification systems assume a natural origin even as ISO detections are about to increase dramatically. The Loeb Scale addresses this gap by providing the first quantitative framework that explicitly includes protocols for evaluating the full spectrum of possibilities, ensuring that future ISOs receive comprehensive assessment regardless of how anomalous their characteristics might be.

**The Need for Comprehensive ISO Classification**

While the vast majority of ISOs will probably prove to be natural objects, any robust classification system must acknowledge that these objects from other star systems could, in principle, encompass a broader range of phenomena. The success of the Torino Scale in classifying near-Earth object impact risks reveals the value of quantitative, graduated classification systems in astronomy (Binzel, 2000). However, the Torino Scale's assumption of natural origin limits its applicability to ISOs, where anomalous characteristics might require more nuanced evaluation.

**The Loeb Scale: Structure and Implementation**

The Loeb Scale (see Table 1) adapts the Torino Scale's proven 0-10 integer framework while introducing fundamental modifications for ISO classification. Where the Torino Scale evaluates only impact probability and kinetic energy for objects of presumed natural origin, the Loeb Scale incorporates multiple observable characteristics including trajectory anomalies, spectroscopic signatures, geometric properties, and other observable characteristics that could distinguish natural from potentially artificial objects. This approach ensures systematic, evidence-based classification as ISO detections increase dramatically in the coming decade.

**Table 1**

*The Loeb Scale Classification Levels*

| Level | Color | Significance Category | Key Observable Criteria |
|-------|-------|----------------------|-------------------------|
| 0 | White | Insignificant | Consistent with known natural phenomena. |
| 1 | Green | Normal Natural Variation | Minor deviations, likely natural variations. |
| 2 | Yellow | Meriting Attention | Non-gravitational acceleration exceeding cometary models. Single major anomaly in trajectory, composition, or morphology. Non-gravitational acceleration is marginally inconsistent with measured outgassing. |
| 3 | Yellow | High Confidence Anomaly | Non-gravitational acceleration vastly exceeding maximum cometary outgassing given absence or weakness of visible coma. Multiple persistent anomalies across observable categories. No satisfactory natural explanation after a comprehensive analysis. |
| 4 | Yellow | Anomaly Meeting Potential Technosignature Criteria | Non-gravitational acceleration exceeding cometary models. Spectral signatures absent in known asteroid taxonomy, including anomalous spectrum inconsistent with solar reflection. Albedo variations inconsistent with known materials. Deviation from Keplerian hyperbolic orbit inconsistent with outgassing models. Unusual shape inferred from lightcurve of reflected |

| | | | sunlight. Trajectory anomalously aligned with planetary orbital planes or selective inner planet targeting. |
|---|---|---|---|
| 5 | Orange | Suspected Passive Technology | Unusual speed. Strong, persistent indicators of artificial, non-operational origin. Surface composition inconsistent with cosmic-ray bombardment for implied age or velocity. Absence of cometary activity despite substantial non-gravitational acceleration. |
| 6 | Orange | Suspected Active Technology | Level 5 criteria plus at least one of the following: (i) Signs of being operational (e.g., maneuvers, signals); (ii) Electromagnetic signals in non-natural origin; (iii) Trajectory changes incompatible with gravitational or outgassing models; (iv) Detection of deployed sub-objects. (iv) Artificial illumination or heat that cannot be explained by solar irradiation. |
| 7 | Orange | Suspected Active Technology with Unclear Intent | Level 6 criteria plus at least one of: (i) Responsive behavior to observations; (ii) Signals of unknown purpose; (iii) Operational intent that cannot be determined or appears potentially hostile. |
| 8 | Red | Confirmed Technology (No Impact) | Direct investigation confirms extraterrestrial artificial origin. No collision trajectory. |
| 9 | Red | Confirmed Technology (Regional Impact) | Confirmed extraterrestrial artificial origin. Impact trajectory with regional consequences. |
| 10 | Red | Confirmed Technology (Global Impact) | Confirmed extraterrestrial artificial origin. Impact trajectory with global terrestrial consequences. |

*Note*. The Loeb Scale extends the Torino Scale to address ISOs, incorporating technological origin assessment. Color coding follows standard risk communication protocols. Level 4 marks the critical threshold where technosignature indicators trigger enhanced observational campaigns. Objects must meet all criteria for Levels 0-3 sequentially. For Levels 4-10, objects may skip levels if higher-level indicators are definitively observed (e.g., electromagnetic signals would justify direct Level 6 classification).

**Levels 0-4**

The Loeb Scale's architecture divides into three zones reflecting increasing deviation from expected natural phenomena. The Green Zone encompasses Levels 0-1, representing objects consistent with known natural processes, though Level 1 allows for minor unexplained variations. The Yellow Zone, covering Levels 2-4, addresses objects with increasingly significant anomalies. Level 2 captures single major deviations, such as trajectory anomalies exceeding gravitational models. Level 3 indicates multiple persistent anomalies, particularly non-gravitational accelerations that significantly exceed cometary outgassing limits.

The critical threshold occurs at Level 4, where technosignature indicators enter formal consideration. This level requires meeting Level 3 criteria plus additional features weakly consistent with artificial origin, such as spectral signatures absent from known asteroid taxonomies, albedo variations inconsistent with natural materials, unusual shape from lightcurve variations, or trajectory alignments with planetary orbital planes having a low probability. Level 4 represents the critical juncture where scientific curiosity must expand to include strategic considerations, the point at which 'What is it?' necessarily becomes 'What does it mean for humanity?' Though remaining within the Yellow Zone as these indicators require verification.

**Levels 5-7**

The Orange Zone encompasses Levels 5-7, marking the qualitative shift to suspected artificial origin requiring immediate strategic response. Level 5 applies when strong persistent indicators suggest artificial but non-operational technology, such as surface compositions inconsistent with expected cosmic ray bombardment or absence of cometary activity despite substantial non-gravitational acceleration. Level 6 elevates the classification when operational signs emerge, including electromagnetic emissions across non-natural frequencies, evidence of propulsion, or detection of deployed sub-probes. Level 7 addresses the complex scenario where technology is confirmed but intent remains unclear or potentially hostile, where detection itself may carry risks.

Unlike Near-Earth Objects (NEOs) that offer mostly years of observational opportunity, the hyperbolic trajectories of ISOs limit observational windows to months or weeks, requiring prepared response protocols. The distinction between natural and technological objects relies on convergent evidence across multiple observables. Following Loeb (2025b), key technosignature indicators include: propulsion signatures causing deviations from gravitational trajectories that exceed cometary outgassing limits; trajectories selectively targeting inner planets or exhibiting improbable alignments with the ecliptic plane; spectral signatures distinguishing artificial illumination or internal heat sources from reflected sunlight; anomalous shapes inferred from lightcurves; surface compositions inconsistent with expected interstellar weathering; electromagnetic signals across artificial frequencies; and evidence of deployed sub-probes. These criteria inform the progression from Level 4 (initial indicators) through Level 8 (confirmed technology).

**Levels 8-10**

The Red Zone, Levels 8-10, manages post-confirmation scenarios differentiated by impact potential. Level 8, confirmed technology posing no impact threat, has no Torino analog but represents humanity's most profound potential discovery. We advise mandatory immediate data release to prevent information hoarding and ensuring discoveries remain the "province of all mankind" per the Outer Space Treaty (United Nations, 1967, Article I). Levels 9 and 10 address regional and global terrestrial impact scenarios respectively, invoking emergency protocols that transcend typical natural disaster response given the added complexity of technological causation, whether from malfunction or intent.

The scale's comprehensive structure serves three essential functions validated by decades of Torino Scale operations. First, definitional precision: when exercises simulate Level 8 scenarios, participants share exact understanding without ambiguity, critical for international coordination. Second, institutional memory: unused upper levels maintain awareness of possibility space, preventing normalization of null results. The absence of Level 10 NEO events fails to recognize the Chicxulub crater; similarly, non-detection of artificial ISOs does not eliminate their logical possibility. Third, capability forcing: Torino's high levels catalyzed billions of dollars in planetary defense infrastructure despite never being invoked. Planetary defense exercises, conducted since 2013, have helped to refine response strategies and inform the operations of NASA's Planetary Defense Coordination Office and missions like the Double Asteroid Redirection Test (DART), while complementing international efforts such as ESA's Space Situational Awareness programme.

Historical precedence reinforces the value of comprehensive frameworks. WHO pandemic levels included Category 5 and 6 (widespread human infection) long before COVID-19, enabling rapid protocol activation when needed (World Health Organization, 2009). Nuclear incident scales extend to Level 7 though only Chernobyl and Fukushima reached this level among thousands of reactor-years (International Atomic Energy Agency, 2008), yet the framework's existence enabled proportionate response when extremes materialized. The Rio Scale for SETI detections similarly extend from 0-10 (Almár & Tarter, 2011), revealing astronomy's comfort with comprehensive possibility frameworks. The Loeb Scale similarly provides common language, clear triggers, and pre-positioned capabilities across the possibility spectrum.

Critics might question including scenarios for which we lack precedent, yet the Torino Scale demonstrates that transparent acknowledgment of full possibility space enhances rather than diminishes credibility. Public acceptance requires honest uncertainty communication, precisely what gradated scales provide. The asymmetry between preparedness costs and potential consequences further justifies completeness. Planetary defense spending (~$200 million annually represents ~1% of space budgets (NASA, 2023) while addressing events with potentially catastrophic consequences. For ISOs potentially harboring evidence of extraterrestrial intelligence, stakes amplify exponentially.

The Loeb Scale represents neither alarmism nor complacency, but measures awareness and preparedness. As the Torino Scale's unused upper levels maintain institutional readiness for rare but consequential impacts, the Loeb Scale ensures humanity possesses a conceptual framework for discoveries that may redefine our cosmic status. That we hope never to invoke Levels 9-10 does not diminish their necessity; preparedness itself may influence outcomes in ways we cannot yet imagine. As detection rates increase from a few ISOs per decade to potentially one ISO per few months, the Loeb Scale ensures each encounter receives systematic evaluation calibrated to its characteristics rather than our preconceptions.

## Case Studies: Applying the Loeb Scale Through Known ISOs

The Loeb Scale's practical utility becomes apparent when applied to the three confirmed ISOs detected to date. These classifications demonstrate how the scale discriminates between natural and anomalous characteristics while maintaining scientific objectivity.

### 1I/'Oumuamua (2017): Level 4 Classification

1I/'Oumuamua represents the most challenging classification case, exhibiting multiple anomalies that accumulate to reach Level 4. Its non-gravitational acceleration of $4.92 \pm 0.16 \times 10^{-6}$ m s$^{-2}$ (Micheli et al., 2018) significantly exceeded limits expected from cometary outgassing, particularly given the absence of detectable coma or carbon-based molecules despite sensitive observations by the Spitzer Space Telescope (Trilling et al 2019). 1I/'Oumuamua's brightness varied by a factor of 10 during its 8-hour rotation period, indicating an extreme geometry with an aspect ratio exceeding 10:1 (Drahus et al., 2018; Meech et al., 2017), that falls outside the distribution of known solar system objects. Its lightcurve suggested a flat disk geometry (Mashchenko 2019), unusual for natural bodies.

These combined anomalies: non-gravitational acceleration without visible outgassing, extreme geometry, and lack of spectroscopic volatiles, meet Level 3 criteria. The additional presence of characteristics weakly consistent with artificial origin (unusual shape, unexplained propulsion) elevates the classification to Level 4, triggering the recommendation for enhanced observational campaigns that, unfortunately, time constraints prevented.

### 2I/Borisov (2019): Level 0 Classification

In stark contrast, Borisov presented as a classical comet despite its interstellar origin (Guzik et al., 2020). Its coma appeared at 2.8 AU from the Sun, consistent with water ice sublimation. Spectroscopic observations revealed CN, $C_2$, and other typical cometary volatiles. The object's morphology, activity pattern, and trajectory followed predictions for a natural icy body. Minor compositional differences from solar system comets (slightly higher $CO/H_2O$ ratio) represent the type of natural variations that place Borisov at Level 0.

### 3I/ATLAS (2025): Level 4 Classification

The third confirmed ISO presents a compelling case for Level 4 classification through an unprecedented accumulation of anomalous characteristics. Most strikingly, its retrograde orbital plane lies within 5 degrees of Earth's ecliptic plane, a configuration with only 0.2% probability for random orientations (Hibberd et al., 2025).

The object's estimated ~20-km diameter creates a profound size paradox: we should have detected a million smaller objects before encountering one of this size[2], and if 3I/ATLAS truly has a ~10 km radius, the implied interstellar mass density would exceed the expected mass budget of ejected rocky materials by 4 orders of magnitude (Loeb, 2025a). This constraint implies either 3I/ATLAS is a comet with a nucleus radius <0.6 km (making its lack of spectroscopic volatiles even more puzzling), or it belongs to an extremely rare population with number density $<5\times10^{-8}$ au$^{-3}$ that somehow favors trajectories toward the inner solar system, adding yet another unlikely coincidence to its profile.

3I/ATLAS displays no spectral features of cometary gas despite observed fuzz preceding the object, distinguishing it from typical comets like 2I/Borisov. Recent observations confirm this paradox: despite weak dust activity (with mass loss rates of only 0.3-6 kg/s) and a visible coma, spectroscopic analysis reveals no cometary gas emissions even at closer heliocentric distances (Santana-Ros et al., 2025). The object exhibits a rotation period of 16.16±0.01 hours with a lightcurve amplitude of ~0.3 magnitude, parameters that could aid in constraining its physical properties. 3I/ATLAS shows progressive reddening in its spectral gradient (from 17.1% to 22.8% per 1000Å over three weeks), suggesting evolving surface or coma composition as it approaches the Sun (Santana-Ros et al., 2025). Finally, its trajectory exhibits remarkable synchronization, approaching unusually close to Venus (0.65 AU), Mars (0.19 AU), and Jupiter (0.36 AU) with a cumulative probability of only 0.005% for random arrival times (Hibberd et al., 2025).

Additional features strengthen the case for Level 4 classification: the object achieves perihelion on the opposite side of the Sun from Earth on October 29, 2025, precisely where a reverse Solar Oberth maneuver would be optimal for spacecraft deceleration while remaining hidden from Earth-based observation. Its approach direction from the Milky Way center hindered early detection, and calculations show that velocity thrusts needed for launches from 3I/ATLAS to intercept inner planets are less than 5 km/s, achievable with conventional propulsion (Hibberd et al., 2025).

---

[2] Loeb, A. 2025, https://avi-loeb.medium.com/welcoming-a-new-interstellar-object-a11pi3z-0b01f1cb4fbc

While Hibberd et al. (2025) explore technological hypotheses for 3I/ATLAS as "largely a pedagogical exercise", they acknowledge the object is most likely natural. Their analysis nonetheless identifies quantifiable anomalies relevant to Loeb Scale classification. These combined anomalies: trajectory alignment, size discrepancy, synchronized planetary approaches, and features optimized for spacecraft operations, meet Level 3 criteria while presenting multiple characteristics weakly consistent with technology, warranting Level 4 classification and triggering recommendations for intensive observational campaigns.

These case studies reveal how the Loeb Scale provides consistent, evidence-based classification while avoiding both excessive skepticism and unwarranted speculation about artificial origins. Importantly, Level 4 classification does not imply artificial origin but rather indicates sufficient anomalies to warrant comprehensive investigation of all possibilities, including technological hypotheses.

**Implementation Through Established Astronomical Governance**

The Loeb Scale's adoption must follow the successful precedent established by the Torino Scale, which the International Astronomical Union (IAU) formally endorsed in 1999. We propose a three-phase implementation pathway: First, the IAU's Division A (Fundamental Astronomy) should establish a dedicated Working Group on ISO Classification. This group, comprising experts in small body astronomy and astrobiology, would refine the scale's technical specifications and validate classification criteria through systematic review of existing ISO data.

Second, following working group recommendations, the scale will undergo community review through IAU Commissions A3 (Fundamental Standards) and A4 (Celestial Mechanics). This process ensures broad scientific consensus while incorporating feedback from observers who will implement the system.

Third, formal adoption will occur through IAU General Assembly resolution, establishing the Loeb Scale as the international standard for ISO classification. The Minor Planet Center will integrate classification protocols into their announcement circulars, ensuring consistent application as new ISOs are discovered. This pathway leverages existing governance structures while maintaining scientific rigor. The Torino Scale's successful twenty-five-year history reveals that graduated classification systems can achieve universal adoption when implemented through established institutional channels.

## Discussion

### Balancing Scientific Rigor with Preparedness Imperatives

The Loeb Scale addresses a critical gap in astronomical infrastructure as we transition from serendipitous ISO discoveries to routine detections. Its primary strength lies in providing the first systematic framework that acknowledges the full spectrum of possibilities without compromising scientific objectivity. By establishing clear, quantitative thresholds for classification, the scale transforms what could devolve into chaotic speculation to structured scientific investigation. The scale's graduated approach offers several advantages over binary natural-versus-artificial determinations. It acknowledges that anomalous characteristics exist along a continuum, allowing for nuanced assessment as observational data accumulates. The explicit inclusion of technosignature protocols at Level 4 and above ensures that potentially significant discoveries receive appropriate scrutiny without triggering premature conclusions.

This balance proves essential given the limited observation windows and data quality constraints inherent in ISO investigations.

However, several observational limitations merit acknowledgment. ISOs' hyperbolic trajectories severely constrain data collection, potentially preventing acquisition of observations necessary for definitive classification. The scale's reliance on multiple convergent indicators may prove challenging when dealing with objects detected late in their solar system passage.

Importantly, as our sample of ISOs remains small, the criteria established here will require refinement as patterns emerge from larger populations. The scale also faces interpretive challenges. The boundary between Levels 3 and 4 represents a particularly sensitive threshold where natural explanations become strained but artificial origin remains speculative. Clear operational guidelines and regular calibration exercises will prove essential to maintain classification consistency across different observer teams and institutions.

Future refinements must incorporate lessons from upcoming detections. As more data is collected, the resulting ISO statistics will enable more precise probability calculations for trajectory anomalies and compositional variations. Machine learning algorithms trained on expanded datasets can assist in rapid preliminary classification, though human judgment must remain paramount for higher-level determinations.

The Loeb Scale's ultimate value extends beyond individual object classification. Like the

Torino Scale before it, its existence shapes institutional preparedness, funding priorities, and technological development. By acknowledging that ISOs might harbor evidence of extraterrestrial technology, however remote that possibility, we ensure our scientific frameworks match the profound nature of potential discoveries. As humanity stands at the threshold of routine ISO detection, the Loeb Scale provides essential infrastructure for systematic investigation. Whether future ISOs reveal only natural astrophysical processes or something more profound, we now possess a conceptual framework ensuring each encounter receives evaluation commensurate with its characteristics. In establishing this scale before the coming deluge of detections, we demonstrate that scientific preparedness can indeed match cosmic opportunity.

## Conclusion

We have presented the Loeb Scale, a newly developed comprehensive classification system that extends the Torino Scale framework to address the unique anomalies of ISOs. By incorporating explicit protocols for evaluating anomalous characteristics and potential technosignatures, the scale provides the astronomical community with a standardized tool precisely when needed most, as the Vera C. Rubin Observatory will increase ISO detection rates by nearly two orders of magnitude.

The scale's practical utility has been demonstrated through classification of all three known ISOs, showing how it distinguishes between mundane and anomalous characteristics while maintaining scientific objectivity. Its graduated 0-10 structure allows for nuanced assessment as observational data accumulates, avoiding premature conclusions while ensuring potentially significant discoveries receive appropriate scrutiny. Implementation through established IAU channels offers a clear path to international adoption, paralleling the Torino Scale's successful integration into planetary defense protocols. As we stand at the threshold of routine ISO detection, the Loeb Scale ensures that each encounter will receive systematic evaluation calibrated to its characteristics rather than our preconceptions. The framework now exists; its success depends on the astronomical community's commitment to implement it before the coming deluge of discoveries.